\documentclass[preprint2,times,tighten]{aastex6}
\usepackage{amsmath,amstext}
\usepackage[figure,figure*]{hypcap}
\usepackage{newtxmath} 
\usepackage[utf8]{inputenc}
\usepackage[T1]{fontenc}

\shorttitle{RQA in astronomical high-contrast imaging}
\shortauthors{M. Stangalini et al.}

\begin{document}

\title{Recurrence quantification analysis as a post-processing technique in adaptive optics high-contrast imaging}
\author{M. Stangalini\altaffilmark{1,2}, G. Li Causi\altaffilmark{3,1}, F. Pedichini\altaffilmark{1,2}, S. Antoniucci\altaffilmark{1,2}, M. Mattioli\altaffilmark{1,2}, J. Christou\altaffilmark{4}, G. Consolini\altaffilmark{3}, D. Hope\altaffilmark{5}, S. M. Jefferies\altaffilmark{6,7}, R. Piazzesi\altaffilmark{1,2}, V. Testa\altaffilmark{1,2}}

\email{marco.stangalini@inaf.it}
\altaffiltext{1}{INAF-OAR National Institute for Astrophysics, Via Frascati 33, 00078 Monte Porzio Catone (RM), Italy}
\altaffiltext{2}{INAF-ADONI, Adaptive Optics National Laboratory, Italy}
\altaffiltext{3}{INAF-IAPS, National Institute for Astrophysics, Via del Fosso del Cavaliere 100, 00133 Rome, Italy}
\altaffiltext{4}{LBTO, University of Arizona, Tucson AZ 85721, USA}
\altaffiltext{5}{Hart Scientific Consulting International LLC, 2555 N. Coyote Dr. 114, Tucson, AZ 85745}
\altaffiltext{6}{Georgia State University, Atlanta, GA 30303, USA}
\altaffiltext{7}{University of Hawaii, Institute for Astronomy, Maui, Hawaii', USA}

\begin{abstract}
In this work we explore the possibility of using Recurrence Quantification Analysis (RQA) in astronomical high-contrast imaging to statistically discriminate the signal of faint objects from speckle noise. To this end, we tested RQA on a sequence of high frame rate (1 kHz) images acquired with the SHARK-VIS forerunner at the Large Binocular Telescope. Our tests show promising results in terms of detection contrasts at angular separations as small as $50$ mas, especially when RQA is applied to a very short sequence of data ($2$ s). These results are discussed in light of possible science applications and with respect to other techniques like, for example, Angular Differential Imaging and Speckle-Free Imaging.
\end{abstract}


\section{Introduction}
Speckle noise and residual stray light resulting from small optical imperfections and seeing-induced aberrations represent one of the major limitations that hamper the detection of faint companions of nearby stars in high-contrast imaging \citep{marois2000efficient, macintosh2005speckle, cavarroc2006fundamental, 2012SPIE.8447E..0BK}. This is also the case in adaptive optics (AO) coronagraphic imaging operating under the best seeing conditions at infrared wavelengths \cite{2013A&A...554A..41M}.\\
To limit the effect of these sources of noise, and increase the detection contrast in astronomical imaging  \citep{hugot2012active}, a number of post-facto techniques have been proposed. Among the many, we recall here the widely used Angular Differential Imaging (ADI), Locally Optimized Combination of Images (LOCI), and Principal Component Analysis \citep[PCA;][]{2006ApJ...641..556M,2007ApJ...660..770L, 2012ApJ...755L..28S, 2012MNRAS.427..948A}. \\
These techniques rely on estimating and subtracting the central PSF together with its quasi-static aberrations from the data, in order to increase the detectability of faint objects in the field of view (FoV). More recently, \citet{2017ApJ...849...85L} proposed an extension of the ADI technique, called Speckle-Free ADI (SFADI), in which residual speckles are identified and numerically masked out in short-exposure images before the latter are piled up just as in the ADI technique. This is possible whenever the exposure time of the images is shorter than the typical lifetime of the speckles themselves, so that they can be identified and efficiently masked out.\\
The same authors successfully tested this technique on the high cadence (1 ms) data acquired at visible wavelengths by the SHARK-VIS forerunner experiment at LBT \citep{2014SPIE.9147E..8FS, 2017AJ....154...74P, 2017JATIS...3b5001S}, demonstrating an increase in contrast of the order of a factor of two over the ADI method. This also demonstrates the advantage of high frame rates that freeze the atmospheric evolution, thus allowing the identification and removal of short-lived speckles from the data.\\
\begin{figure*}[!ht]
   \centering
   \includegraphics[width=18cm, clip]{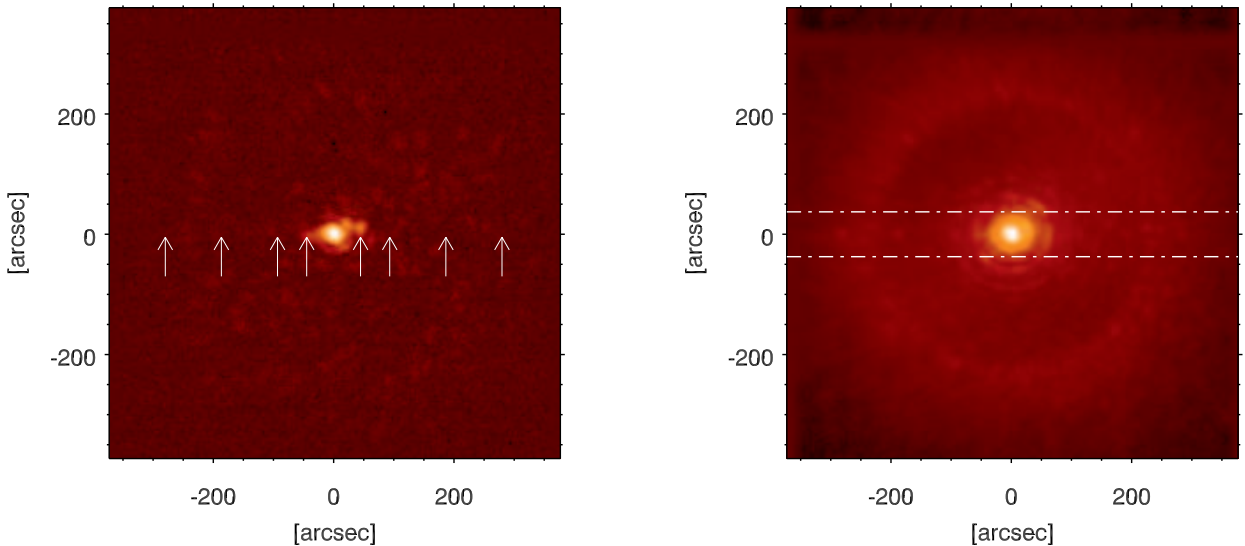}
   \caption{Left: short exposure (1 ms) image of the target GLIESE 777. Right: sum of 2000 images equivalent to 2 s integration time of the same target. The arrows in the left panel indicate the position of the injected synthetic planets, while the dashed lines in the right panel highlight the region used for the tests of the RQA.}
    \label{imgs}
   \end{figure*}

In this regard, forthcoming fast high-contrast imagers, like SHARK-VIS, will be able to deliver, at least for a subset of bright sources, high cadence data sets that will be made of several thousand, or even millions, of images.\\
However, even if these techniques make good use of the high acquisition rate, the temporal information in the data sequence is not fully exploited.\\
Many authors have proposed alternative approaches to the techniques based on PSF subtraction: these alternative approaches rely on the statistical discrimination of the planet signal with respect to speckle statistics \citep[see for instance][to mention a few]{1995A&A...298..544L, Canales99, 1999SPIE.3866..165C, 1999ApOpt..38..766C, 2001OptEn..40.2690C, 2004EAS....12...89A, 2006ApJ...637..541F, 2010JOSAA..27A..64G, Frazin16}. More in particular, it was shown \citep{1999ApOpt..38..766C} that the probability density function (PDF) of intensity fluctuations in AO corrected images can be modeled as a Rician distribution: this evolves from an exponential to a Gaussian as one moves from the halo of speckles towards the core of the PSF. If there is a planet buried in the noise, the corresponding pixel signal is the superposition of a random process (i.e. the AO residuals), a low-frequency drift due to non-common path aberrations (NCPAs), and the planet signal itself. If there is enough information in the temporal structure of the data, one can separate the statistical processes and obtain an unambiguous detection of the faint source. \\
The key point here is that kHz-rate imaging freezes atmospheric evolution so that the resulting datasets contain additional temporal information that can be exploited for faint sources detection. This information is completely lost in standard long exposure observations, so that fast imaging provides the additional possibility of applying statistical techniques which are completely separate from PSF subtraction methods like ADI or PCA and, as such, are complementary to these and can offer an independent validation of the results.\\
Here we test the ability of the Recurrence Quantification Analysis \citep[RQA;][]{marwan2003encounters}, a well know technique of non-linear time-series analysis in the phase space, to identify changes in the underlying intensity statistics in astronomical images where faint sources are buried in noise.

   \begin{figure}[!ht]
   \centering
   \includegraphics[width=7cm, clip, trim=0mm 0mm 0mm 0mm]{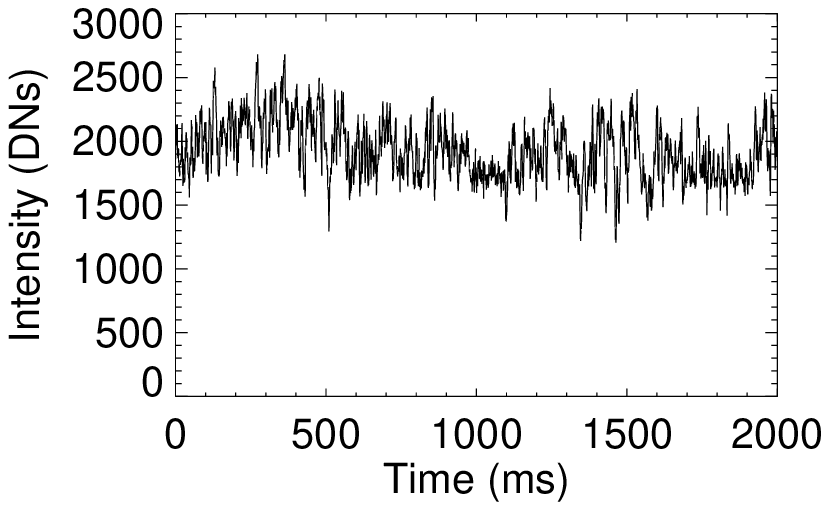}
   \includegraphics[width=7cm, clip, trim=-20mm 0mm 20mm 0mm]{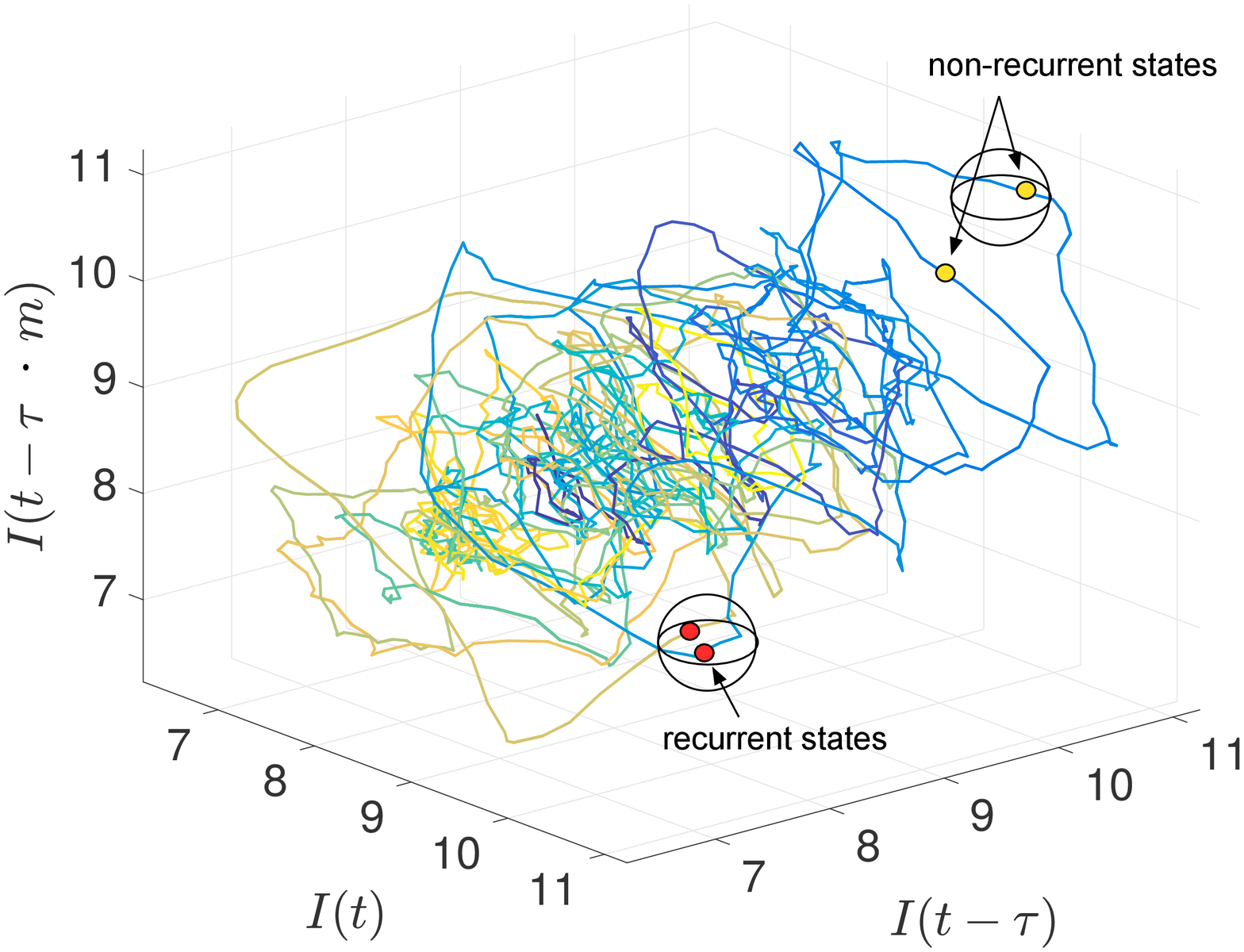}
   \includegraphics[width=7cm, clip, trim=0mm 0mm 0mm 0mm]{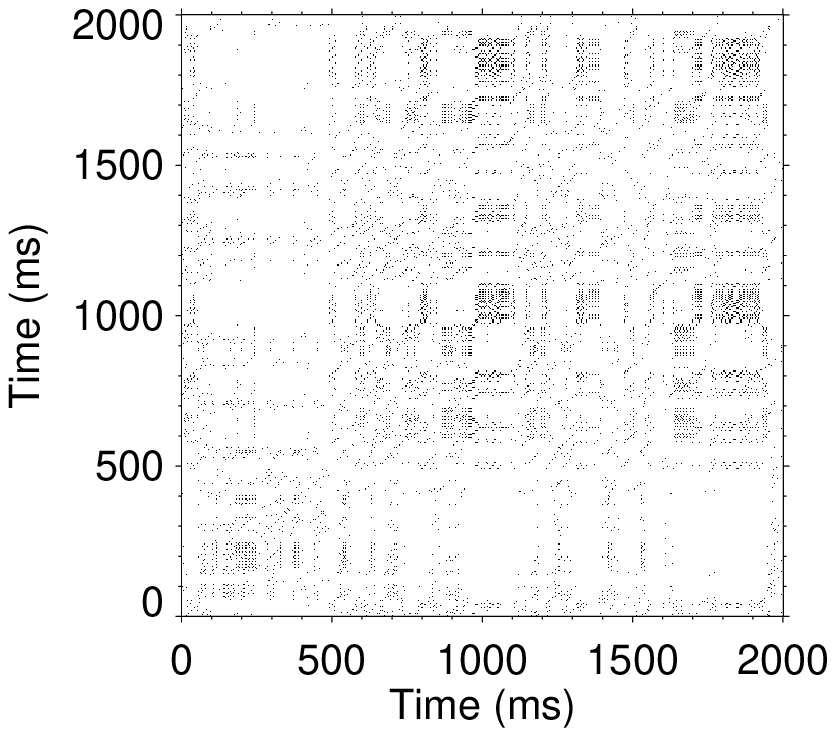}
   \caption{Top: Intensity time sequence at the center of the star peak. Middle: phase space reconstruction obtained with the time-delay technique for $m=3$ and $\tau=10$ ms. Here we highlight, for example, two recurrent states in which the trajectories visit the same region of the phase space at different times. The colors encode the temporal evolution from dark blue to yellow. Bottom: thresholded RP obtained from the phase space reconstruction.}
    \label{RPs}
   \end{figure}   
   
The detection of small signals embedded in noise is a general problem that affects not only astrophysics and high-contrast imaging but also other research fields (e.g. cryptography and denoising of radar signals, to mention a few). It has been shown that the detection of small signals in exceptionally noisy environments, by exploiting variations of the underlying statistics of the process caused by the presence of the small signal itself, can be successfully accomplished using RQA \citep[][]{zbilut2000recurrence, Rohde_stochasticanalysis, michalowicz2008signal, marwan2009comment, iwanski1998recurrence, marwan2003encounters}.

   \begin{figure*}[!ht]
   \centering
   \includegraphics[width=6cm, clip, trim=0mm 0mm 0mm 0mm]{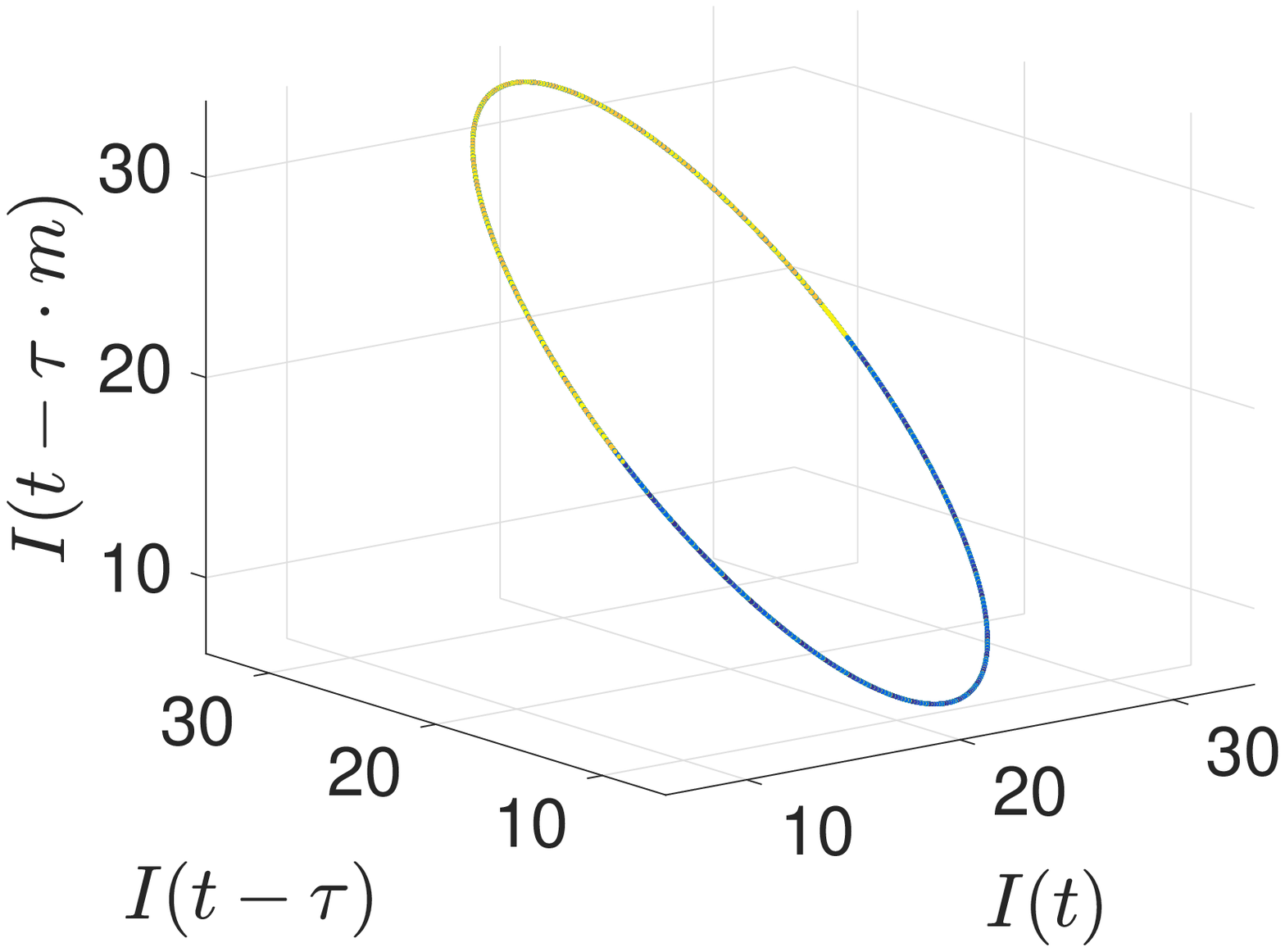}
   \includegraphics[width=7cm, clip, trim=0mm 0mm 0mm 0mm]{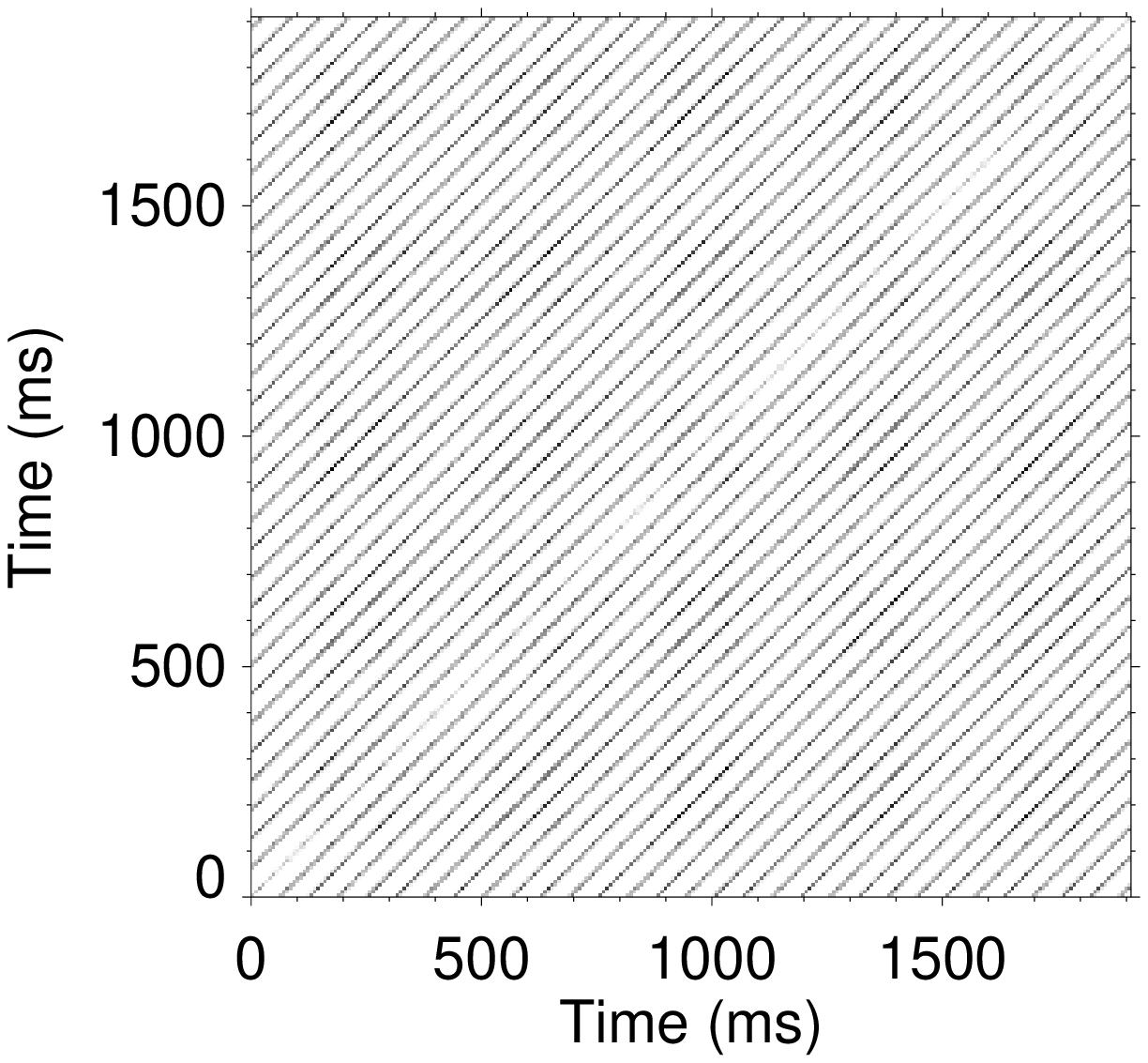}
   \includegraphics[width=6cm, clip, trim=0mm 0mm 0mm 0mm]{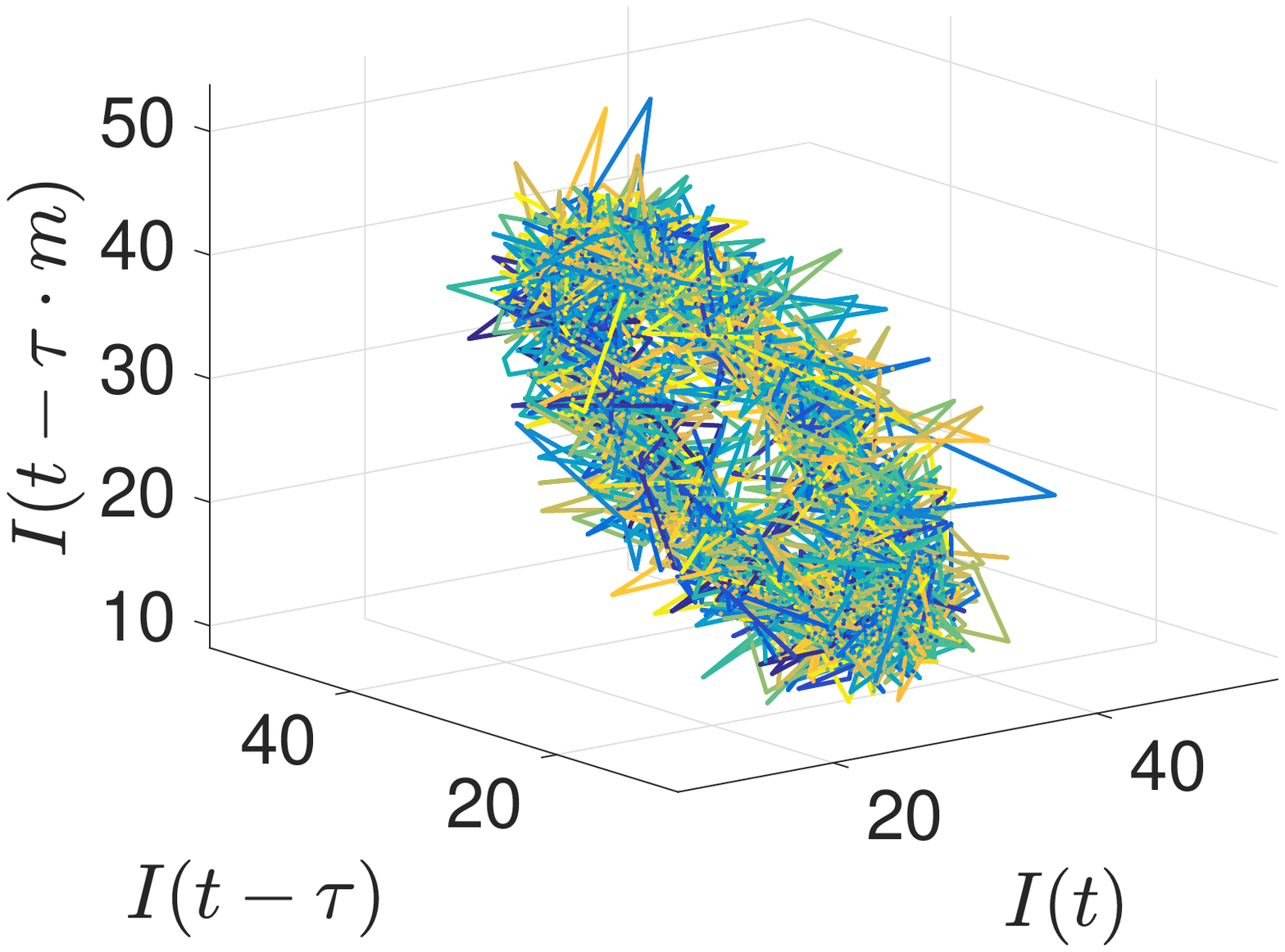}
   \includegraphics[width=7cm, clip, trim=0mm 0mm 0mm 0mm]{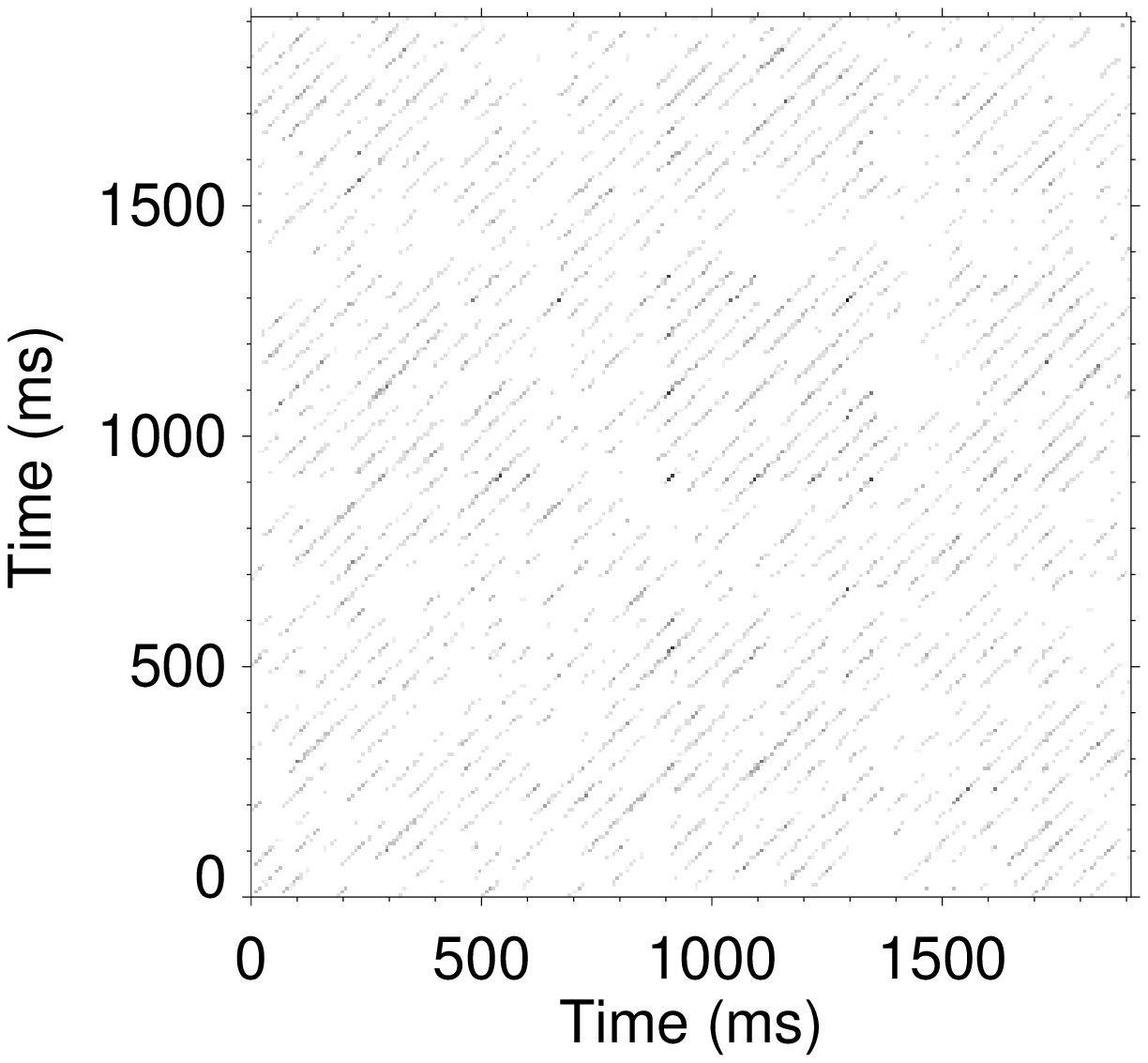}
   \caption{Upper panels: phase space reconstruction of a sinusoidally modulated signal ($m=10$, $\tau=10$ ms; left), and its RP (right). Lower panels: Phase space reconstruction of a sinusoidally modulated signal with a planet with a contrast of $10^{-3}$ superimposed. The planet in injected with a Poisson statistics (left) and its RP (right). Embedding dimensions and RP thresholds are the same for all cases to facilitate their comparison.}
    \label{toy_models_sin}
   \end{figure*}   
   
      \begin{figure*}[]
   \centering
   \includegraphics[width=6cm, clip, trim=0mm 0mm 0mm 0mm]{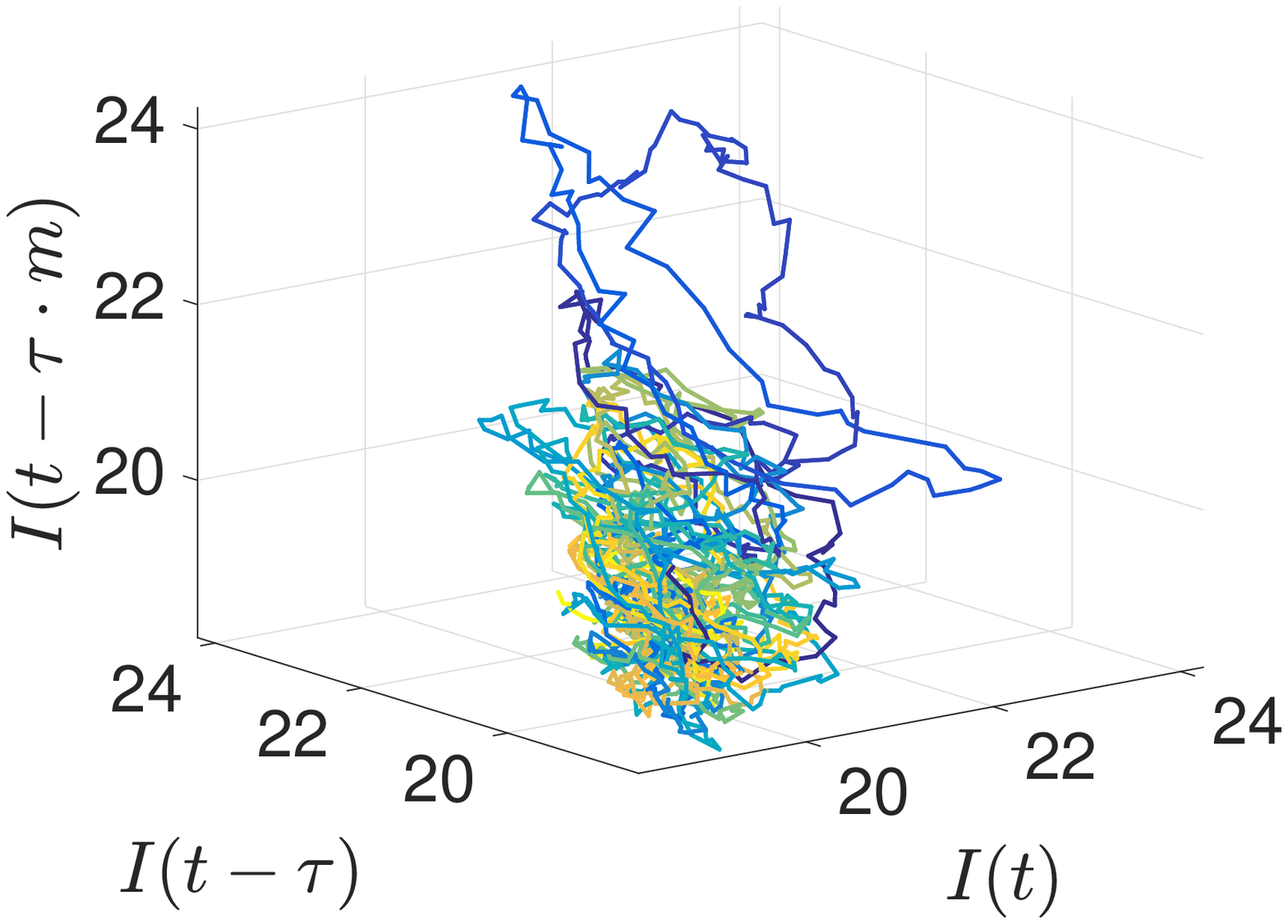}
   \includegraphics[width=7cm, clip, trim=0mm 0mm 0mm 0mm]{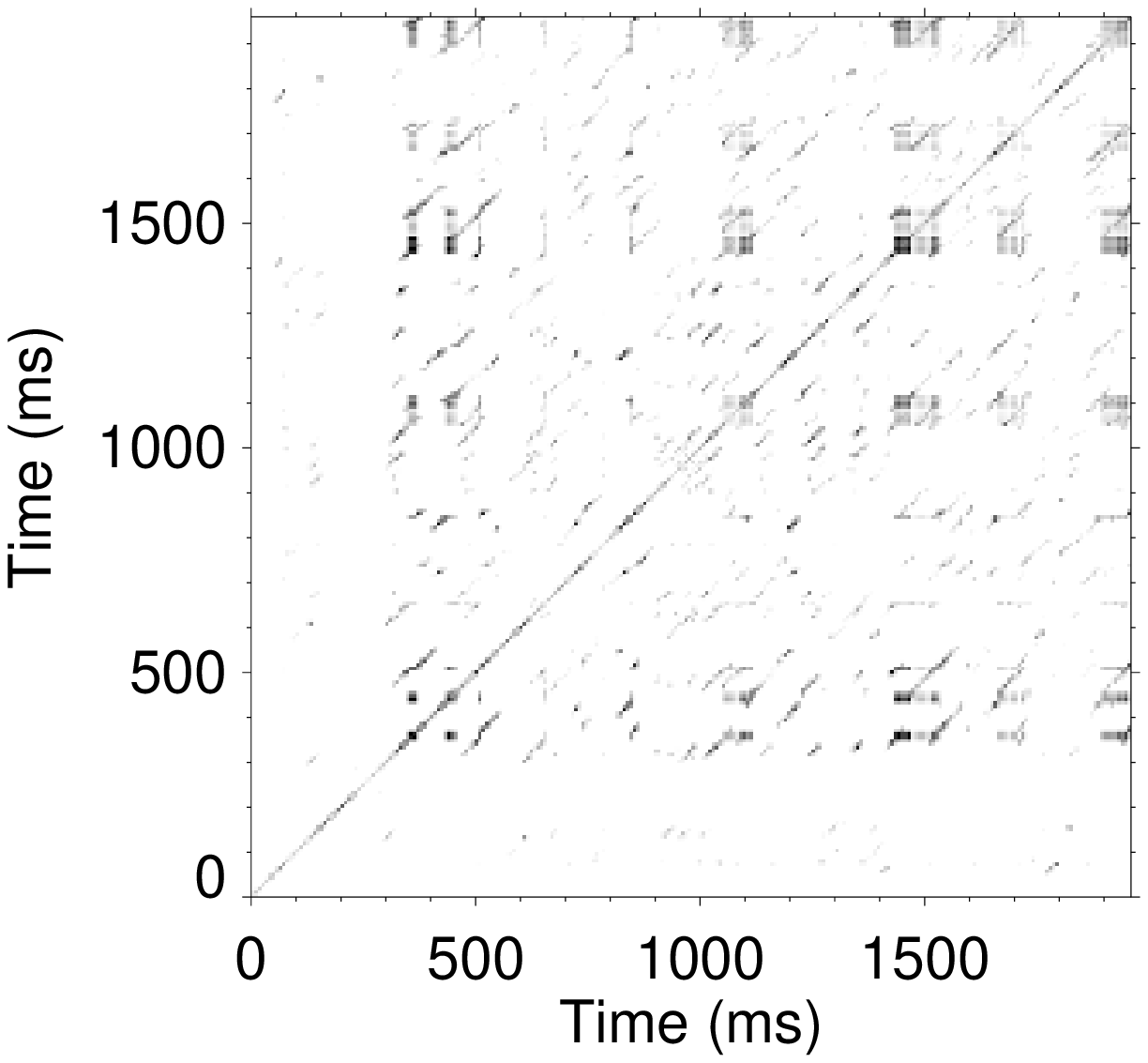}
   \includegraphics[width=6cm, clip, trim=0mm 0mm 0mm 0mm]{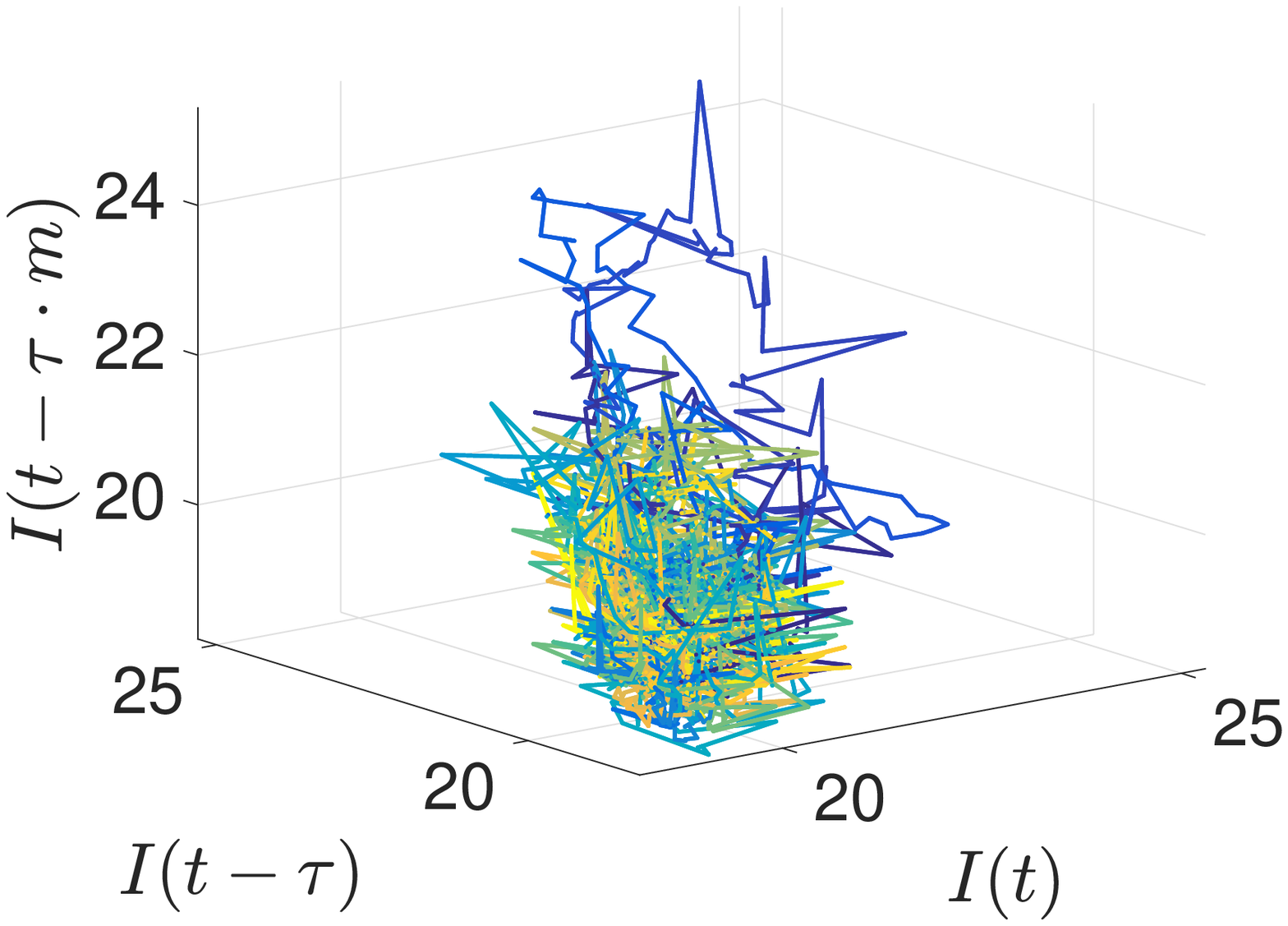}  
   \includegraphics[width=7cm, clip, trim=0mm 0mm 0mm 0mm]{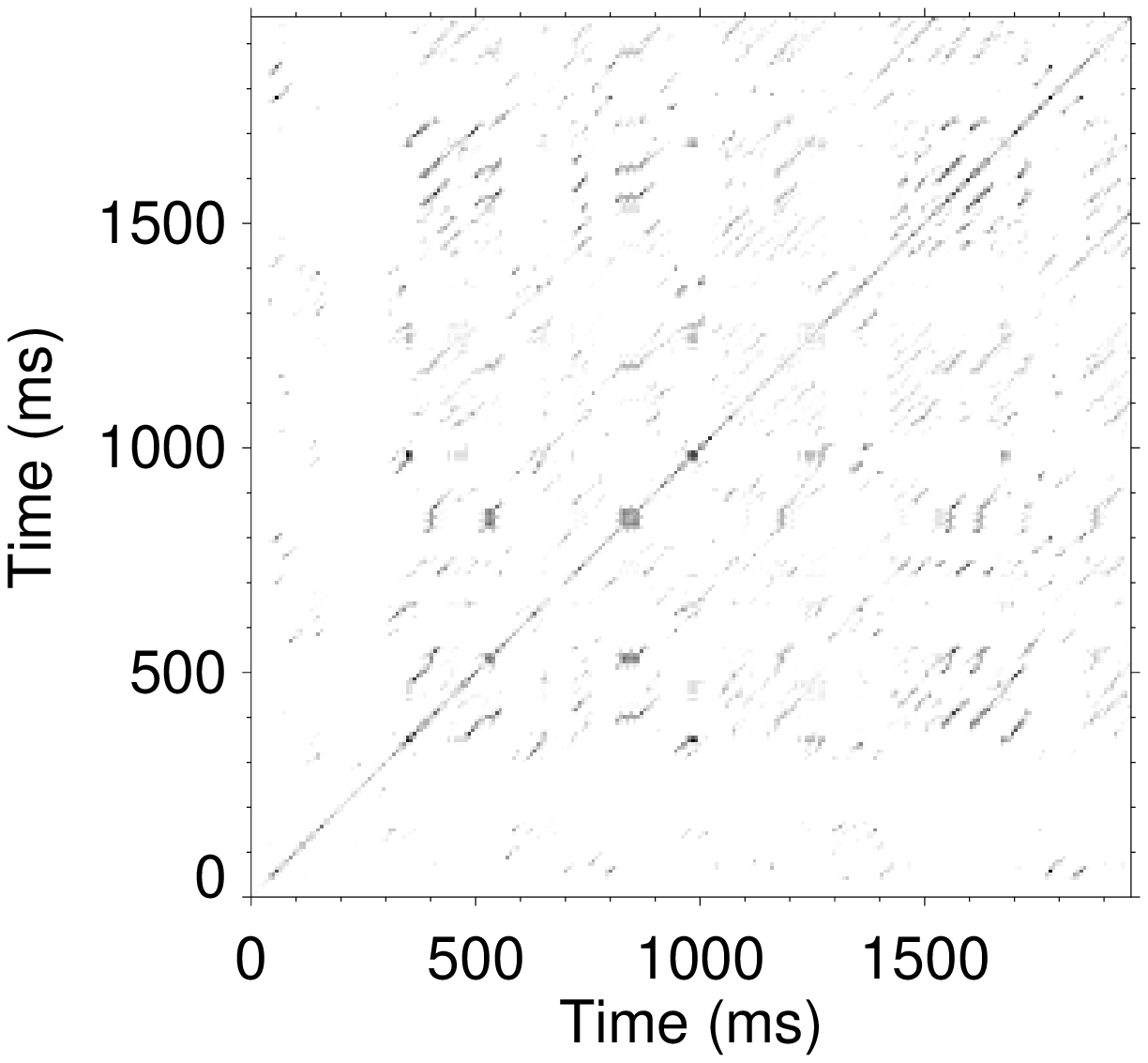}
   \caption{Upper panels: phase space projections of a modulated signal with the same spectrum of the AO residuals ($m=10$, $\tau=10$ ms; left), and its RP (right). Lower panels: Same but with a planet with contrast of $10^{-4}$ superimposed. Embedding dimensions and RP thresholds are the same for all cases to facilitate their comparison.}
    \label{toy_models_modulation}
   \end{figure*}  
   
The RQA is based on the fact that the recurrence of states in the phase space is a general property of dynamical systems. Through the quantification of these recurrent states in the phase space, RQA offers a practical way to examine different complexity indicators in a dynamical system \citep[see for example][for more details]{thiel2004much, webber2005recurrence, zbilut2006recurrence, marwan2007recurrence}. These measures can be exploited to reveal small changes in the dynamics associated with the presence of a small signal embedded into exceptionally noisy environments.\\
Although this technique has already been applied to signal detection in other contexts, as far as we know it has never been applied before to astronomical high-contrast imaging. This work represents a first attempt to explore the possibility of using RQA for the identification of faint astronomical sources embedded into speckle noise and provide a first assessment of its performances. To this aim, we test the RQA on a sequence of fast cadence images acquired by the SHARK-VIS forerunner experiment at visible wavelengths.  
 
\section{Data set}
The data set used in this work consists of a series of 1 ms exposure images of the star Gliese 777 (1 ms cadence), acquired with the SHARK-VIS forerunner experiment \citep{2017AJ....154...74P} at the Large Binocular Telescope (LBT) in June, 2015 (see left panel of Fig. \ref{imgs}).
The experiment consisted in a series of short test observations between February and June 2015 to verify the LBT AO system performance at visible wavelengths ($600 - 900 $nm). 
The experimental setup was minimal and composed of only two optical elements before the detector: one divergent lens to get a super sampling of the PSF (point spread function) and a $40$ nm FWHM filter centered at 630 nm. 
The pixel scale was set at $3.73$ mas and the camera employed was a Zyla sCMOS camera manufactured by Andor Inc\footnote{http://www.andor.com/}. The total duration of the data series taken into account is $20$ min (or equivalently 1\,200\,000 images), although here we only make use of a subsample of 2000 images (2 s). During the frame acquisition, the LBTI AO system \citep{esposito2010first} was correcting 500 modes in closed loop, and seeing was in the range $0.9 - 1.2$ arcsec.\\
NCPAs were reduced by means of the procedure described in \cite{2015aoel.confE..36E} down to $80$ nm rms.\\
The fast exposure time adopted for the observations allows us to freeze the evolution of atmospheric speckles and to easily recover the residual jitter in the focal plane.
For additional details we refer the reader to \cite{2017AJ....154...74P} and \cite{2017JATIS...3b5001S}. We remark that the forerunner experiment is a pathfinder of the SHARK-VIS high-contrast imager, now approved and at the time of writing in its construction phase.\\
The data calibration process consisted in the dark frame subtraction and image registration with sub pixel accuracy.\\ In the left panel of Fig. \ref{imgs}, we show a short (1 ms) exposure image of the target, while in the right panel of the same figure we show the image obtained by summing up 2000 re-centered frames, which is equivalent to a $2$ s exposure time.

\section{Methods}
Recurrence states, regions of the phase space which are visited recurrently, represent a general property of dynamical systems \citep{kac1947notion, iwanski1998recurrence}. Indeed, the way the trajectories of a dynamical system explore the phase space itself provides a unique characterization of the process being investigated. For example, random processes, in contrast to deterministic ones, have the tendency to maximize the explored volume of the phase space and usually lack recurrent structures. In order to facilitate the study of the recurrence structure of a dynamical system represented by a particular time series, \citet{eckmann1987recurrence} introduced a specific diagram, called a recurrence plot (RP), that represents in a 2D domain the recurrences of the physical process in the phase space.
The starting point of this procedure is the phase space representation of the time series under investigation (the time series of the intensity fluctuations in a pixel in our case). This is generally referred to as the \textit{embedding procedure} and can be obtained, for example, through the time-delay technique \citep{zbilut1992embeddings}. In this case, the vectors that constitute the phase space representation of the system are constructed from time delayed measurements of the scalar time series.
In more detail, our time-discrete measurement 
$I(t_{i})=I(i \Delta t)$, where $i=1,...,N$ and $\Delta t$ is the sampling time, can be represented through the time-delay technique as a vector \textbf{x} in an $m$-dimensional phase space 
defined as:
\begin{equation}
\textbf{x}(t_{i})= \sum_{j=1}^{m} I(t_{i+(j-1)\tau}) \textbf{e}_{j},
\end{equation}
where $m$ is the embedding dimension, $\tau$ the chosen time delay, and $\textbf{e}_{i}$ are the base vectors of an $m$-dimensional orthogonal coordinate system.\\

In Fig. \ref{RPs} (middle) we show an example of this time-delay embedding of the time-series corresponding to the intensity fluctuations at the peak of the star in our dataset, which is shown in the upper panel of the same figure. In the same plot, we also illustrate an example of recurrent and non-recurrent states. The RP is a time-time plot in which each point (i,j) is shaded according to the distance between the points $X_{i}$ and $X_{j}$ on the trajectory in the phase space. The closeness of the states of the system at different times (recurrences) determines specific features of the RP, such as clusters of points, which provide information on the nature of the investigated dynamical system.\\
In the phase space we can set a threshold that defines a distance within which two states can be considered recurrent. This corresponds to a neighborhood radius in the phase space (see middle panel of Fig. \ref{RPs}). 
The bottom panel of figure \ref{RPs} shows the RP derived from the time series of the intensity fluctuations shown in the upper panel of the same figure after application of the threshold. The convention adopted is such that a recurrent state in the phase space is marked by a black dot in the RPs. The threshold is chosen on the basis of the specific application. In our case we selected a threshold equal to $1$ and this value comes from tests aimed at the optimization of the final contrast.\\
Before discussing in more detail the information that can be extracted from RPs, it is useful to illustrate a few simple cases to qualitatively show how RPs are affected by the small signal we may expect from a companion planet. In order to do so, we have extracted from our dataset the image of a real speckle and modulated it in time with a sinusoidal function and with the same average amplitude spectrum of oscillations measured in the data. In the latter case, the amplitude spectrum is extracted as an average of the spectra in a $10 \times 10$ pixel box. This results in a speckle with a random modulation but with the same spectrum of oscillations measured on-sky. We then added a synthetic planet with a Poisson statistic in the simulations box to study how the phase space is perturbed with respect to the case without planets and how these information are translated into the RPs. 

\subsection{Phase space representation and RP of a periodic system.}
In Fig. \ref{toy_models_sin} (upper left panel) we show the phase space projection of the signal of a speckle modulated in time with a sinusoidal function (without planet), obtained for an embedding $m=10$ and $\tau=10$. Note that such a large value of $m$ allows a better visualization of the perturbation to the trajectories in the phase space. The simulated time series consists of 2000 samples. The trajectories in the phase space describe a recurrent pattern and all the states are visited recursively (i.e. recurrent states). In the upper right panel of the same figure we show the RP of this system. The RP displays diagonally aligned lines that indicate a stationary evolution of the system. Here we note that the spacing between the diagonals corresponds to the main period of the system.\\
In the lower panels of the same figure, we also show the phase space reconstruction and RP of the same statistical process, after injecting a faint planet whose contrast is $10^{-3}$ with respect to the central PSF of the data from which the image of the speckle is extracted. Here we see that the photons of the faint planet result in a perturbation of the trajectories of the dynamical system in the phase space. The signal is the superposition of two processes, a periodic one (i.e. the sinusoidally modulated speckle) and the time series associated with the arrival of the planet photons, which have a Poisson distribution. The perturbations of the trajectories due to the planet counts make the explored phase space larger with respect to the purely periodic case. As a consequence of this, the RP too shows changes due to the superposition of the two statistical processes. In particular, we observe a disruption of the diagonal lines which represent the jumps seen in the phase space with respect to a smooth evolution of the trajectory. It is worth noting here that, due to the much smoother character of the sinusoidal function with respect to the planet signal, the phase space reconstruction dramatically changes despite the small contrast of the planet itself. This highlights the advantages of this technique and such data visualization. However, it is also worth noting that, as we will see in the next section, in the presence of noise and a non-periodic signal it might not be so easy to disentangle the two processes by simply looking at the phase space reconstructions. In addition, in this specific case, the amplitude of the sinusoidal function was chosen in such a way to improve the visualization of the phase space reconstruction.

\subsection{Phase space representation and RP of a random system.}
The same analysis described in the previous section was performed by randomly modulating the speckle intensity with an amplitude spectrum taken directly from the temporal evolution of the intensity fluctuations in our data. As already said, in this case the speckle was modulated by using the average amplitude spectrum extracted from a box of $10 \time 10$ pixels in the data. In the upper left panel of Fig. \ref{toy_models_modulation} we show the phase space representation of the dynamical system associated with the intensity fluctuations at the center of the speckle. The RP of this dynamical system is shown in the upper right panel of the same figure and displays a number of disruptions, broken diagonal lines and other over-densities of recurrence states that highlight the temporal evolution of the dynamics seen in the phase space (e.g. non constant density of recurrent states, isolated points in the trajectories). In the lower panels of the same figure we show the same plots after injecting a planet with a contrast of $10^{-4}$. In this case too, the superposition of the two statistical processes determines a perturbation of the trajectories in the phase space. This results in a change of the RP which, although similar, shows different traits such as, for example, the disappearance of overdense regions of recurrent states (see for example $t=1500$ ms). Even in this case, the perturbation of the trajectories due to the planet increases the explored phase space volume and this results in a decrease of the overall density of the recurrent states as highlighted by the overall decrease of the black dots in the RP. Please note that the RPs shown here are obtained using the same threshold.\\

These two simplistic cases show that the recurrence structure of a RP defines the dynamical system that is investigated. In the following list we summarize the main features of a RP and their physical interpretation:
\begin{itemize}
\item homogeneity -  stationary process;
\item fading of the recurrences in the upper left and lower right corners - signal contains drifts;
\item disruptions - non-stationary behaviour / abrupt transitions in the dynamics;
\item isolated points - heavy fluctuations in the process, uncorrelated behaviour;
\item diagonal lines - similar evolution at different times typical of deterministic systems;
\item vertical lines - some state changes slowly with time; trapping (i.e. laminar systems).
\end{itemize}

The recurrence structure of a RP can therefore be exploited to investigate the dynamical system and represents a unique signature from which one can infer important information about its inner nature. From the simple simulations we analysed it is clear that the counts from the planet perturb the trajectory of the system in the phase space and this is reflected by the changes of the RP structures.
From the RP one can gather important information about the dynamical system and a possible change of the trajectories in the phase space resulting from the superposition of different processes (i.e. the signal of the planet). It is worth noting that the information content that can be extracted from the recurrence structure is maximized for short integration (a few ms) images. In this regard, fast frame imaging offers the possibility to apply additional detection methods exploiting the temporal structure of the dynamics.\\  
A way to quantify this information relies on the distribution of points that form specific features. For instance, one can measure the overall density of recurrent points in the RP (i.e. the recurrence rate; RR), the fraction of recurrent states which form diagonal lines (i.e. a measure of the determinism of the process; DET), and the fraction of points along vertical lines that can be interpreted as a measure of laminarity (LAM). This is precisely what the RQA technique does \citep[see for example][]{zbilut1992embeddings,trulla1996recurrence, thiel2004much, webber2005recurrence}. \\
Indeed, the RQA is based on the analysis of the distributions of recurrence points in the vertical and diagonal lines of RPs \citep[for a review see e.g.][]{marwan2003encounters, marwan2007recurrence}. The technique also allows the study of other complexity indicators of dynamical systems, although these are not useful for our goal \citep[for a review see e.g.][]{marwan2007recurrence, zbilut2006recurrence}. \\
Here we restrict our attention to the density of recurrent states RR, the determinism DET, and the laminarity LAM. We investigate the time series of intensity fluctuations at each pixel in the FoV, with the aim of highlighting possible changes in the RQA parameters that can be ascribed to the presence of a faint source and, in particular, to the perturbation of the trajectories of the dynamics in the phase space. With reference to the RPs show in Fig. \ref{toy_models_sin} and \ref{toy_models_modulation}, it is already possible to visually see that, with the planet, the overall density of recurrent points (i.e. RR) is smaller than the case without the planet, and this reflects the increase of the volume explored by the system in the phase space and caused by the perturbations to the trajectories. In order to compare the RQA measures at different locations we normalized the time series of intensity fluctuations by their standard deviation before performing the analysis. This is a good practice when comparing different statistical processes \citep[see for instance][]{marwan2009comment}.\\
In this work we used the command-line recurrence plots code, which is part of the TOCSY (Toolboxes for Complex Systems) toolbox\footnote{The command-line recurrence plots code is freely available at the following link: http://tocsy.pik-potsdam.de/} to estimate the RQA measures. This code has two steps. The first one consists in the phase space reconstruction through the time delay embedding method described by Eq. (1), the second one consists in the construction of the RP with a user specified distance threshold, and in the measure of the density of recurrent points aligned in the vertical/horizontal and diagonal direction. More specifically, the RR, DET, and LAM are defined as:
   \begin{figure*}[!ht]
   \centering
   \includegraphics[width=18cm, clip, trim=5mm 15mm 0mm 20mm]{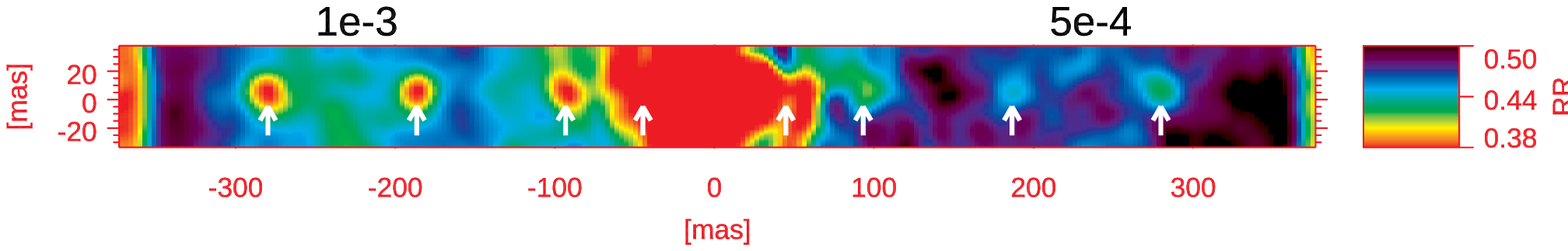}   
   \includegraphics[width=18cm, clip, trim=5mm 15mm 0mm 25mm]{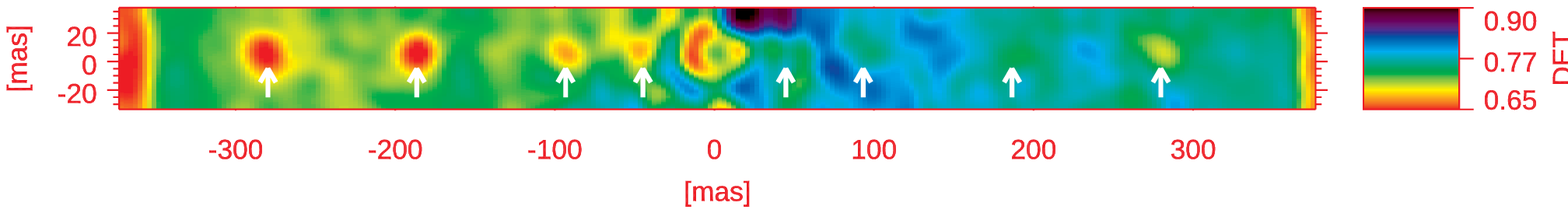}
   \includegraphics[width=18cm, clip, trim=5mm 15mm 0mm 25mm]{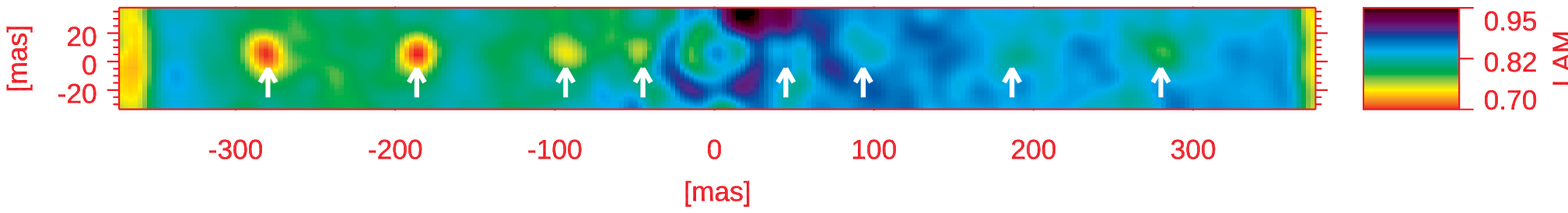}
    \includegraphics[width=18cm, clip, trim=5mm 5mm 0mm 25mm]{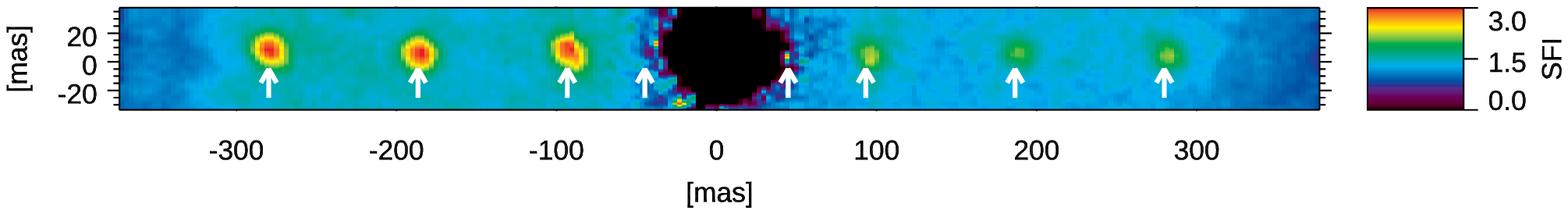}
   \caption{From top to bottom: RR, DET, LAM, SFI maps. The maps are obtained by processing 2000 sequential images.  In the case of RQA maps, embedding parameters are $m=1$ and $\tau=10$.}
    \label{RQA}
   \end{figure*}      
   
\begin{equation}
RR=\frac{1}{N^{2}} \sum_{i,j,=1}^{N} R_{i,j},
\end{equation}
\begin{equation}
DET=\frac{\sum_{l=l_{min}}^{N} l P(l)}   {\sum_{1}^{N} l P(l)},
\end{equation}
\begin{equation}
LAM=\frac{\sum_{v=v_{min}}^{N} v P(v)}   {\sum_{1}^{N} v P(v)},
\end{equation}
where $N$ is the total number of points, $R_{i,j}$ the RP, $P(l)$ the histogram of the lengths $l$ of the diagonal lines, and $P(v)$ the histogram of the lengths $v$ of the vertical lines. $l_{min}=3$ and $v_{min}=3$ in our case. For more information about the algorithms and methods employed in the toolbox (time-delay embedding, generation of RPs, and RQA application) we refer the reader to \cite{marwan2007recurrence}.
To estimate the performances of this technique in the detection of faint astronomical objects embedded into noise, we injected synthetic sources into the real images, by rescaling the PSF of the central object and adding photon noise with a Poisson statistics. This approach for the assessment of a post-processing technique in high-contrast imaging has already been employed by \citet{2012MNRAS.427..948A} and later by \citet{2017AJ....154...74P} and \citet{2017arXiv170903181L} on the same data set.\\
The white arrows in Fig. \ref{imgs} (left panel) indicate the positions of the injected sources. In particular, the planets on the left-hand side of the central object are injected with a contrast ratio of $10^{-3}$, while those on the right-hand side have a contrast of $5 \times 10^{-4}$. \\
It is important to note that, given the limited flux of the central source in ms images (i.e. $1500-2000$ ADU; see upper left panel in Fig. \ref{RPs}), a contrast of the order of $5 \times 10^{-4}$ implies that photons from the faint source at a particular pixel position will not be present in all the images of the sequence.
   
\section{Results}
\begin{figure}[]
   \centering
   \includegraphics[width=9cm, clip, trim=1.5cm 0.5cm 0cm 0cm]{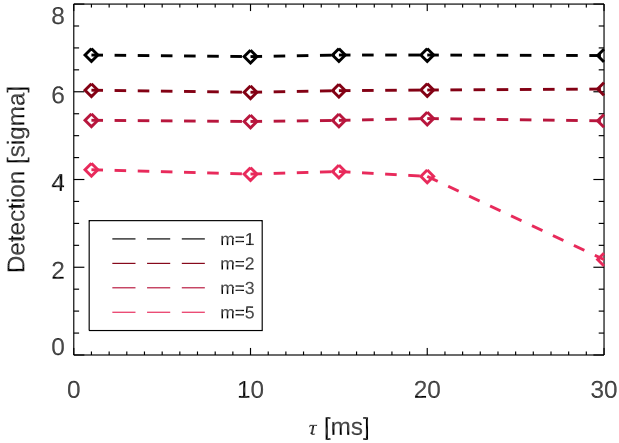}  
   \caption{Average detection as a function of $\tau$, in units of standard deviation, for different values of $m$. The detection curves are obtained by averaging the signals corresponding to the $10^{-3}$ sources in apertures with a radius equivalent to the FWHM of the PSF centered on the planet. The detection contrast is obtained by considering an aperture of $4.5$ pixels, and is performed after subtracting the image background through a $360\deg$ angular median and a spatial median filter with a $22$ pixel radius.}
    \label{tradeoff}
   \end{figure}

In Fig. \ref{RQA}, we show the maps of the estimated RQA parameters RR, DET, and LAM, obtained by applying the technique to a portion of the data sequence as short as $2000$ images, or equivalently $2$ s. Here we zoom on the region marked by the rectangular box in Fig. \ref{imgs}.\\
Here we used the embedding parameters $\tau=10$ ms and $m=1$. The time delay parameter is chosen on the basis of the results of the mutual information analysis in \citet{2017JATIS...3b5001S}, where it is shown that the de-correlation time of the speckles in the same data set is on average $10$ ms. The choice of the $m$ parameter will be clarified below.\\
Both sets of injected faint sources with $10^{-3}$ and $5 \times 10^{-4}$ contrasts, respectively, are visible in all maps of Fig. \ref{RQA}, despite the short duration of the data series used, with the only exception of the sources located at $50$ mas from the central object, which are only seen in LAM and DET maps for a contrast of $10^{-3}$. The detection contrasts will be estimated and discussed in Sect. 4.2, where the RQA maps are compared to the SFI technique. It is worth noting here that, in the $2$ s integrated image in the right panel of Fig. \ref{imgs}, only two of the four objects with contrast $10^{-3}$ are (barely) visible, and only at separation angles larger than $150$ mas. The RQA maps, on the other hand, do show a distinct  change in the statistics in correspondence with the injected sources. This means that the RQA identifies a change of the underlying dynamics of the process at the pixel locations where there is a superposition of two statistical processes; namely the time series of the photons of the faint source, and the speckle noise.\\
RR, DET and LAM all display a decrease of their value at the positions of the planets, although we note that the best contrast is obtained in the RR. Please note that the color table of RQA maps was reversed to facilitate the comparison with the SFI map (see below).\\
This simultaneous decrease of DET and LAM may appear contradictory. Indeed, since these two measures represent the level of determinism and laminarity of the system, one may expect them to be anti-correlated. However, \citet{schinkel2008selection} noted that the DET RQA measure does not relate exactly to the mathematical notion of the term "determinism", but rather underlines the fact that deterministic processes have usually a larger number of diagonal lines in RPs, if compared to purely stochastic processes. Furthermore, \citet{marwan2013recurrence} have also argued that, in some physical systems, the increase of the measure of DET together with that of LAM can be understood as a slowing down of the dynamics, typical of tipping points. For this reason, one should be careful in interpreting, for example, a change of DET or LAM as a change of the determinism or laminarity of the system.\\ 
Nevertheless, we are not interested in the precise physical understanding of the dynamical systems governing the intensity fluctuations in a system, but rather in using the RQA parameters to identify a change in the dynamics that reflects the presence of a faint source embedded into the speckle noise. Our results show that RQA is a very promising tool in this sense.\\
The choice of the parameters of the embedding dimension $m$ is crucial for the optimization of the detection of planets (i.e. for the statistical discrimination of the process associated with the faint source dynamics). The optimal set of parameters may depend on the particular case. For this reason, we have made a trade-off analysis where we have studied the planet's signal-to-noise ratio as a function of the two embedding parameters $m$ and $\tau$. In more detail, at each set of embedding parameters we have measured the planet's detection contrast, defined as the encircled signal at each planet's position in the RR map, divided by the rms of the encircled signal at any other pixel positions. The results of this test are shown in  Fig. \ref{tradeoff}, where we plot this detection contrast as a function of $\tau$, for different values of the embedding dimension $m$.\\
Here we see that the best results are obtained for $m=1$, independently of $\tau$. However, in order to be on the safe side, in agreement with \cite{2017JATIS...3b5001S}, we have chosen $\tau=10$ ms, which is the de-correlation time scale of the speckles. 

\subsection{SFI-RQA comparison}
In this section we compare the RQA with the speckle-free imaging (SFI) method \citep{2017ApJ...849...85L}. This method is a 
particular case of SFADI where the angular rotation of the images is not performed.

SFI differs from SFADI in that it does not subtract a median inter-speckles background, in order to avoid source self-subtraction when field rotation is very small, as in this case of a 2 s acquisition duration.

\cite{2017ApJ...849...85L} have shown that SFADI and SFI result in a better detection contrast with respect to that achieved with a conventional ADI technique (which in any case would not be applicable to this short sequence.\\
In order to test and compare the two techniques in the same conditions, here we have applied the SFI method on exactly the same data series (2000 short exposure images) used to test RQA. The SFI speckle detection threshold was optimized for the detection of sources at radial distance smaller than $100$ mas. \\
The resulting SFI detection map is shown in the bottom panel of Fig. \ref{RQA}. Here we can see that all objects at or beyond $100$ mas are visible.  In this regard, we note that, despite SFI results in a larger detection contrast, RQA allows the detection of objects much closer to the central star ($R< 100$ mas). Indeed in this region the SFI speckle identification and their subsequent suppression is hampered by the increased intensity of the inner core of the PSF, resulting in a lower performance of the SFI algorithm. This is one of the advantages of RQA over SFI however, another interesting aspect relies in the possibility of combining the two techniques. Indeed, as we will see in the next section, taking advantage of the weak correlation of the residuals between SFI and RQA maps, this can result in a significant increase of the contrast.
\begin{figure}[!h]
   \centering
   \includegraphics[width=9cm, clip, trim=20mm 25mm 0mm 0mm]{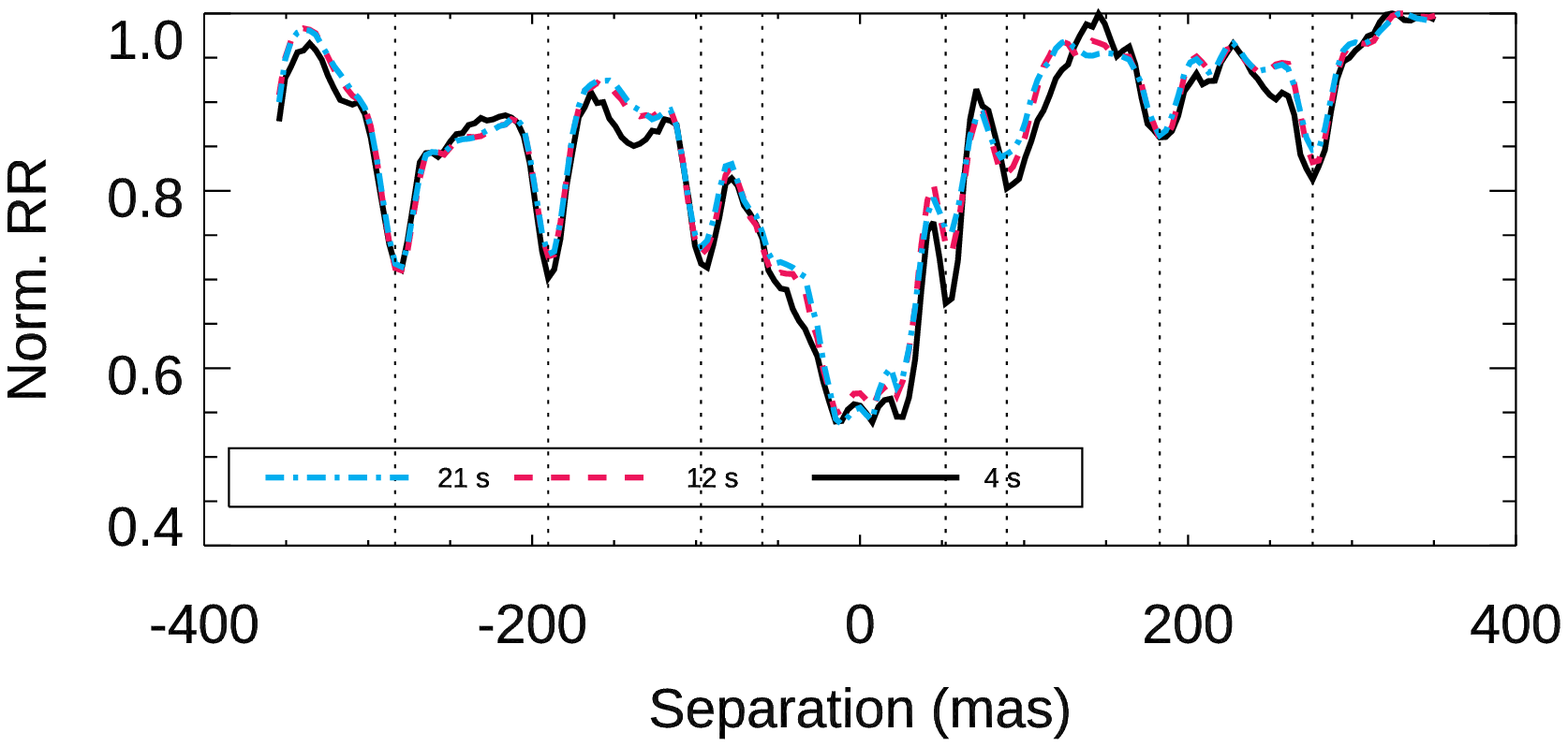}   
   \includegraphics[width=9cm, clip, trim=20mm 25mm 0mm 10mm]{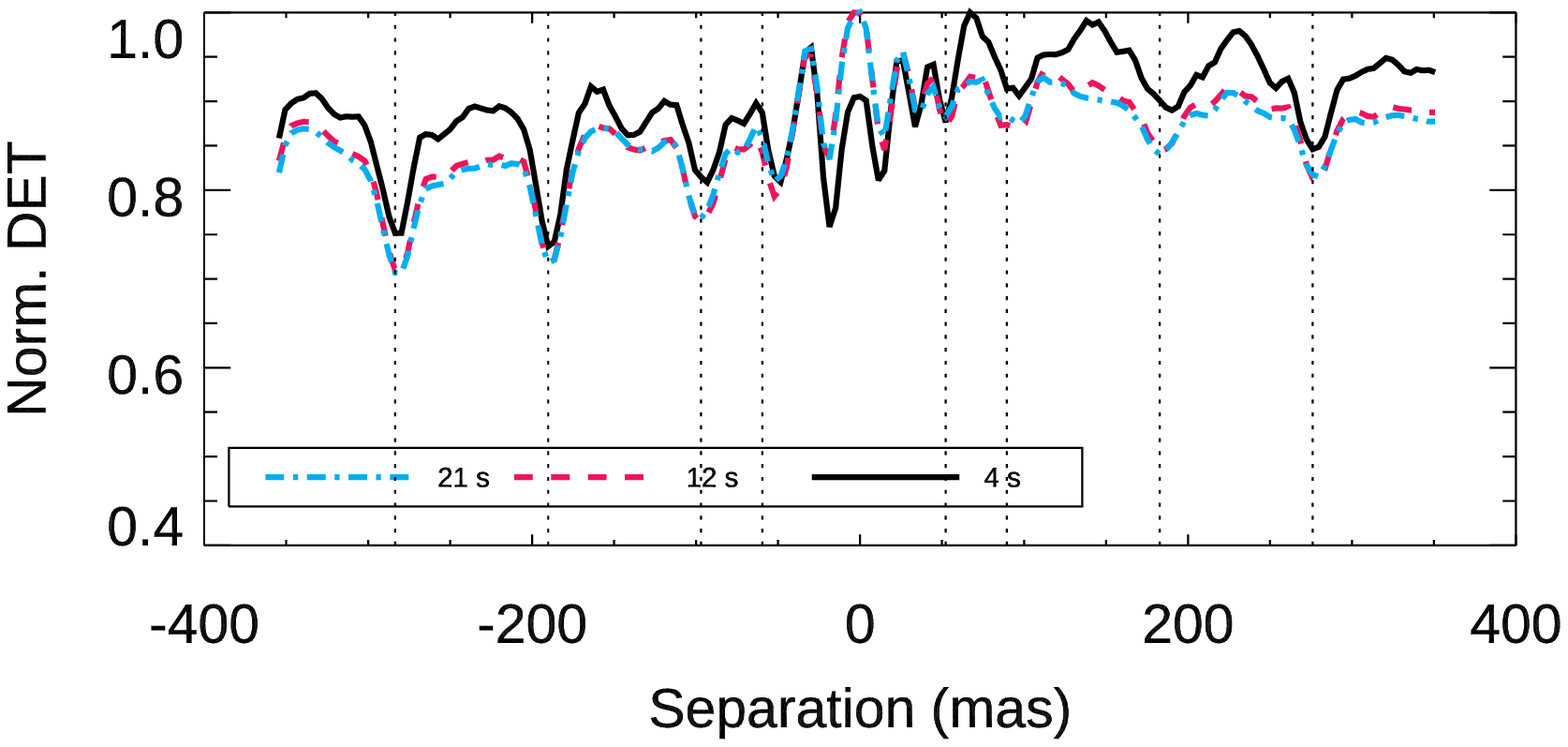}
   \includegraphics[width=9cm, clip, trim=20mm 0mm 0mm 10mm]{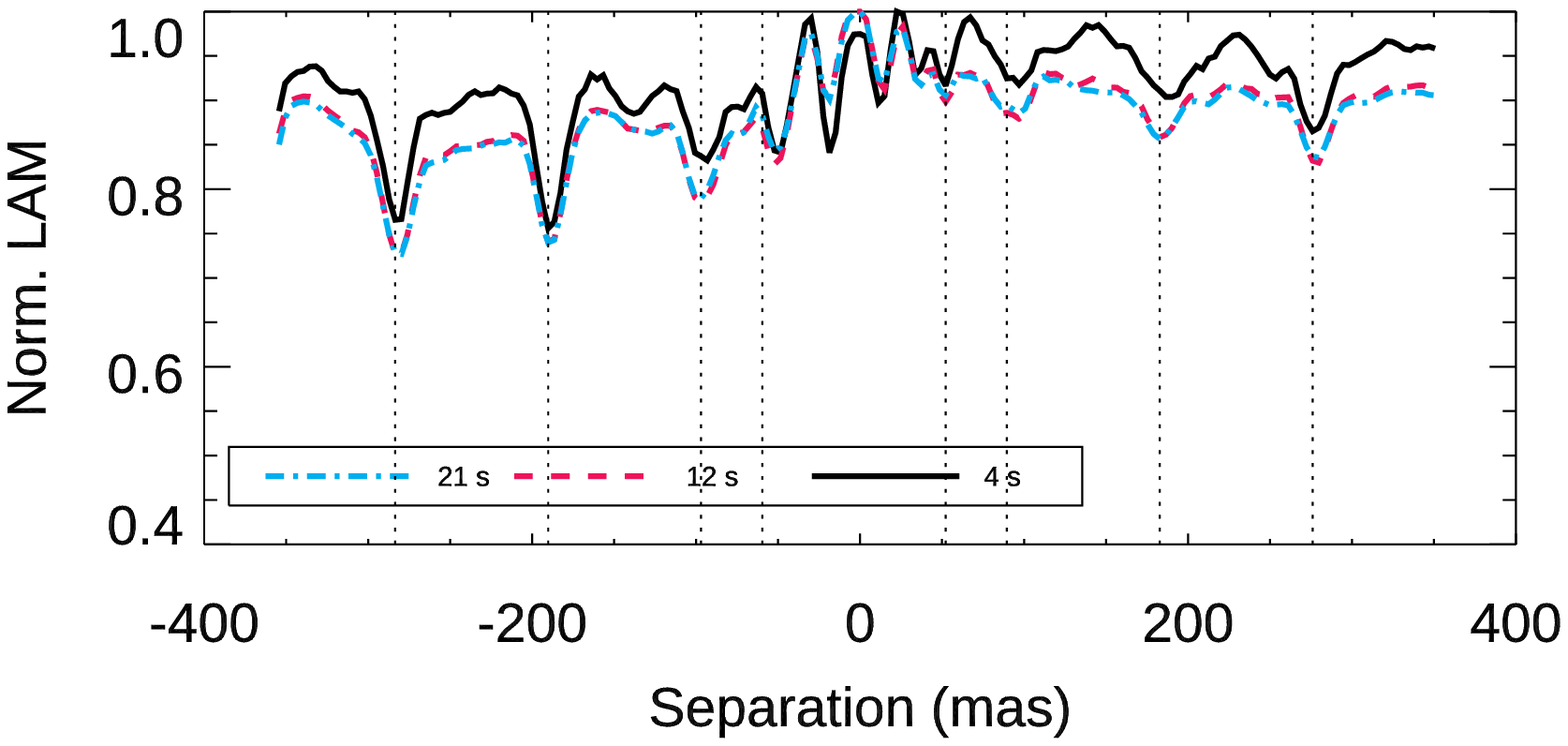}
   \caption{From top to bottom: RR, DET, LAM in a cut passing through the central object, for different lengths of the data sequence used in the analysis (i.e. $2$ s, $12$ s, and $21$ s).}
    \label{integration_effects}
   \end{figure}      

\subsection{Detection contrast}
In table \ref{table:detect} we report the detection contrasts estimated for the RQA parameters, SFI, and their combination. To compute these contrasts for each of the five maps, we used the same procedure adopted for the tradeoff analysis of the embedding parameters in Sect. 4, where we have already shown the contrast ratio, in units of $\sigma$, as a function of the embedding itself. We measured the encircled integral value with an aperture  radius of 4.5 pixels (equals to the FWHM of the PSF) centered on each planet (the signal S). Then we divided this by the r.m.s. of the same measurements performed on every background pixel (the noise N) in apertures that fit in the annulus at the radial distance of each injected planet. This is done after removing the radially averaged background from each map. Hereafter, we will refer to the background subtracted SFI map by adding a superscript in the notation (e.g. $SFI^{*}$).  This allows the resulting S/N to be directly compared to the false detection probability. However, since as in our case (see Fig. \ref{distributions}) the distribution of residuals might not be gaussian \citep[see for instance][]{2006ApJ...641..556M, 2017ApJ...842...14R}, we have computed the significance relative to $3 \sigma$ directly from the cumulative distributions of the residuals themselves, and this is $97.7\%$ in all cases analysed here.
   \begin{figure*}[!ht]
   \centering
   \includegraphics[width=5.8cm, clip, trim=5mm 0mm 0mm 8mm]{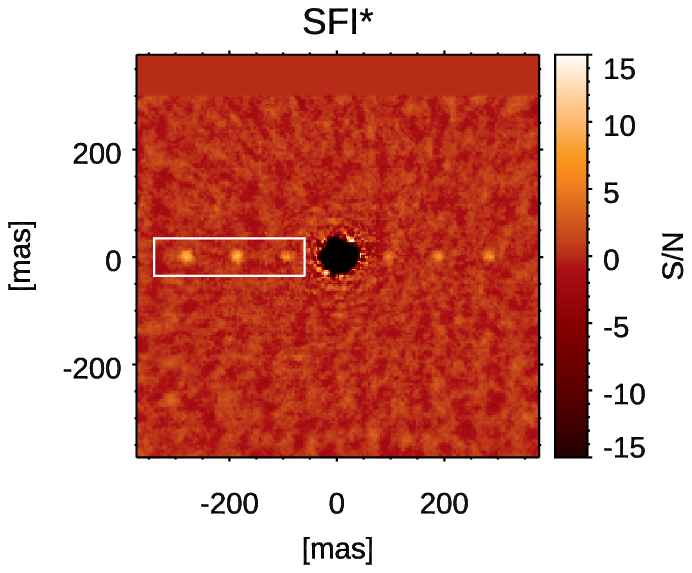} 
   \includegraphics[width=5.8cm, clip, trim=5mm 0mm 2mm 15mm]{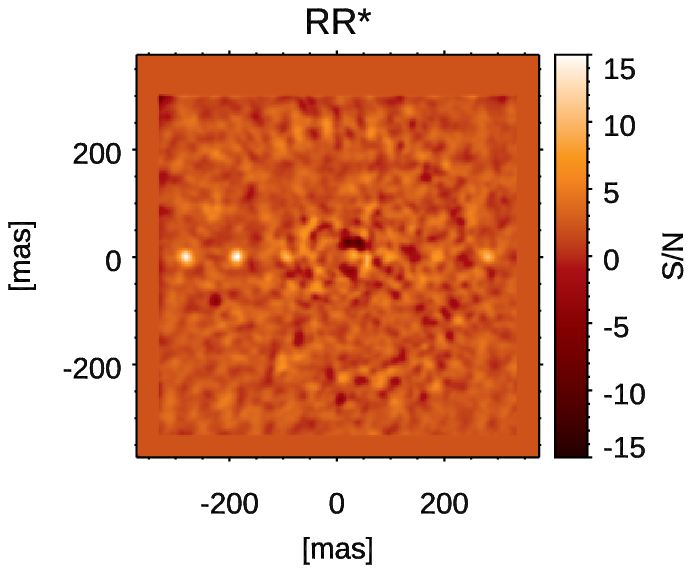} 
   \includegraphics[width=5.8cm, clip, trim=5mm 0mm 0mm 8mm]{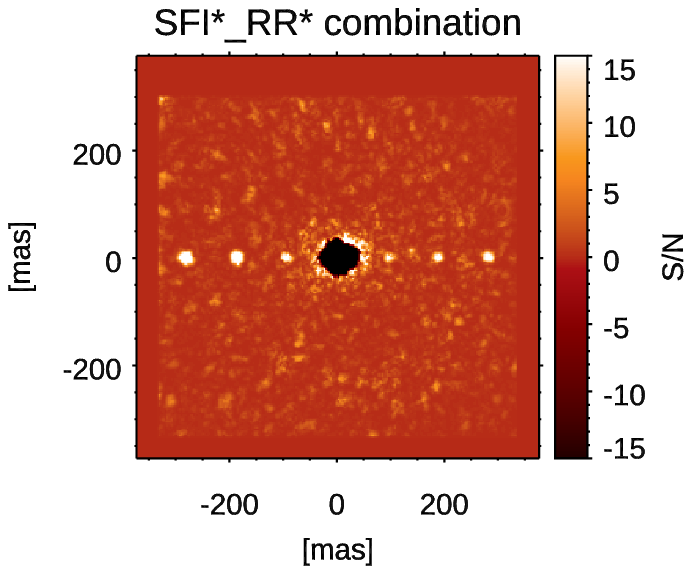} 
   \caption{Full FOV background subtracted SFI* (left), RR* (center) maps and their combination (right). In order to facilitate the comparison, the maps are normalized to the $\sigma$ computed in the region outside the planets.}
    \label{correlation}
   \end{figure*}
The detection contrasts are rather similar for the different RQA measures, however we note that the combination of the RQA and SFI technique does result in a significant improvement of the detectability.\\
The combination of the two techniques is done by multiplying for example the quantity $(1-DET)$ by the SFI residual intensity: $(1-DET) \times SFI$. The quantity $(1-DET)$  increases at the spatial locations of the planets, while it decreases elsewhere. This is done for all RQA parameters. Similar to SFI, hereafter we will refer to these background subtracted quantities as to $DET^{*}$, $RR^{*}$, and $LAM^{*}$. Both RQA parameters and the SFI intensity map mutually increase at the locations where the planets are located, and this allows their combination.\\
In Fig. \ref{correlation}, we show the $SFI^{*}$, $RR^{*}$, and $SFI^{*}$-$RR^{*}$ maps used to estimate the detection contrast. In order to facilitate the comparison, the maps are normalized to the $\sigma$ in the region outside the strip containing the planets. Here we see, in agreement with the detection contrasts in Tab. \ref{table:detect}, a significant increase of the contrast when the two techniques are combined.\\
In order to understand this enhancement, it is useful to analyze the the probability density functions (PDFs) of the residuals from the two techniques and their combination. These are shown in Fig. \ref{distributions} where we plot the PDFs of the residuals of $SFI^{*}$, $RR^{*}$, and their combination, in the area marked by the white box in the left panel of Fig. \ref{correlation}. These PDFs are computed after the subtraction of the mean value and the normalization to the standard deviations. This allows a direct comparisons of the three PDFs themselves. In the PDFs we observe two populations, a low intensity population of pixels due to the residual noise, mostly speckle noise, and a high intensity population which represents the planets.  It is worth noting that the distribution of intensities of the combined $SFI^{*}\_RR^{*}$ map is non-linearly stretched by a factor of $\sim 1.4$ with respect to the single PDFs of $SFI^{*}$ and $RR^{*}$ maps. As a result, the high intensity population of the pixels with the planets is much better detached from the peak due to residual noise, thus increasing the contrast. An important parameter to be considered is the kurtosis of the distributions. This measures the weight of the tails of the distribution and goes from $6.1$ and $4.5$ for $SFI^{*}$ and $RR^{*}$ respectively, to $28.8$ for their combination, and this results from the non-linear stretching of the distribution. It is also useful to note that the correlation coefficient between $SFI^{*}$ and $RR^{*}$ maps computed outside the horizontal strip where the planets are located is $0.4$, while in the regions containing the planets is $\sim 0.9$. This is an important aspect for understanding the physical mechanisms responsible for this increase of the contrast when combining the two techniques. Indeed, since the two methods make use of two completely independent approaches (i.e. speckle removal and statistics of intensity fluctuations), the residuals, in contrast to planets, are only weakly correlated. This means that, the combination of the two technique acts like a sort of "matched filter", with the filter built upon the RQA method. This increases the overall contrast of the most correlated part of the maps (i.e. the planets).\\
   \begin{figure}[!ht]
   \centering
   \includegraphics[width=8cm, clip, trim=5mm 0mm 0mm 8mm]{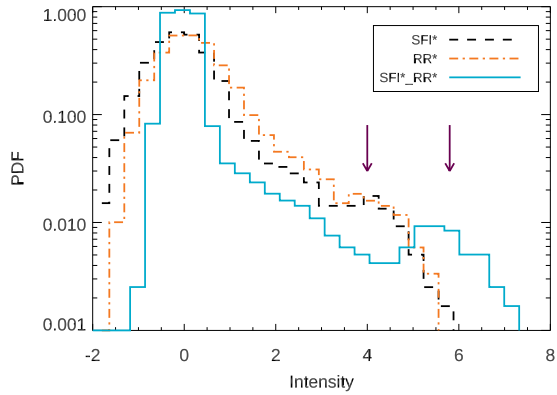} 
   \caption{PDF of the intensity residuals in the box shown in the left panel of Fig. \ref{correlation} for $SFI^{*}$, $RR^{*}$, and their combination. The vertical arrows highlight the position of the planet samples and the stretch of the distribution after combining $SFI^{*}$ and $RR^{*}$.}
    \label{distributions}
   \end{figure}

\subsection{Effects of the data length}
\begin{figure}[!ht]
   \centering
   \includegraphics[width=8cm, clip, trim=0mm 0mm 0mm 0mm]{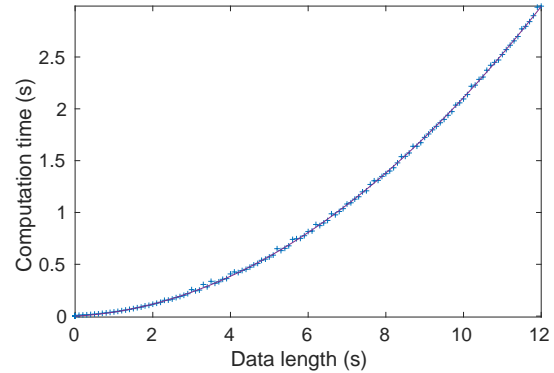}   
   \caption{Wall-clock computation time for a single pixel as a function of the length of the data series. The test was conducted on a Intel i7-4790K CPU @ 4.00GHz, with solid state disks on a single core.}
    \label{computation_time}
   \end{figure}

In addition to the above analysis, we have also studied the effect of the data length on the RQA parameters. In order to limit the computational time we restrained ourselves to a $21$ s data series.\\
In Fig. \ref{integration_effects} it is possible to see the results of this analysis. In particular, we show the normalized RQA parameters in a cut passing through the central star. It appears that extending the lengths of the data series used, at least at these contrast ratios and up to a maximum length equivalent to $21$ s, has no significant impact on the final contrast of the RQA maps. However, although the RQA parameters of a particular process (e.g. the determinism) should be an intrinsic feature of the process itself, in practice their estimation, and therefore the signal-to-noise ratio of the RQA detection maps, can depend on the length of the data series. \\
The application of the RQA to longer sequences implies a significant increase of the computation load. In Fig. \ref{computation_time} we show the wall-clock computation time as a function of the data length for one pixel. The benchmark was conducted on a Intel $i7-4790K$ CPU @ $4.00$GHz, with solid state disks and using a single core. After fitting the data with a second-order polynomial, we extrapolated the computational time needed to process a 20 minute data series (i.e. the length of the SHARK forerunner data). This amounts to $7.5$ hours/cpu. Considering the FoV of the high frame rate mode of SHARK-VIS (i.e. $200 \times 200$ pixels), we estimate a total computation time of $300000$ hours/cpu. This means almost two months on a typical cluster with $200$ nodes. For the application of RQA to longer time series and the estimate of its contrast limits we refer the reader to a future work on the subject.\\
   
\section{Discussions and conclusions}
In this work we have explored the possibility of using RQA for the detection of faint sources in astronomical images. Indeed, \citet{zbilut2000recurrence, marwan2009comment} have shown that RQA can be exploited to detect small signals embedded into exceptionally noisy environments. The main idea behind this is based on the capability of RQA to identify small changes of the underlying dynamics, and act as a statistical discriminator of the intensity fluctuations due to the faint source from the noise it is embedded in.\\

\begin{table}[]

\centering
\begin{tabular}{c c c c}
\hline\hline
Case & S/N $10^{-3}$ planets $[\sigma]$  & S/N $5 \times 10^{-4}$ planets  $[\sigma]$ \\ [0.5ex] 
\hline
RR*& 6.7 & 3.6 \\
DET*& 7.3 & 2.9 \\
LAM*& 8.3 & 3.4  \\
SFI* & 7.6 & 5.0 \\
DET*-SFI* & 64.0 & 15.6 \\
LAM*-SFI* & 72.5 & 17.5 \\
RR*-SFI* & 56.6 & 18.3 \\ [1ex]
\hline
\end{tabular}
\label{table:detect}
\caption{Average S/N, in units of $\sigma$, of the different categories of planets injected for RQA measures, SFI, and a combination of both methods.}
\end{table}

In this regard, it is worth recalling that \citet{2010JOSAA..27A..64G} have shown that faint sources in long sequences of images can be successfully identified by the change of the underlying statistics in the pixels occupied by the sources themselves. In detail, the parameters of a Rician distribution modeling the probability density function of the intensity fluctuations at the source locations are different with respect to those corresponding to another pixel. This piece of information can be exploited to identify the source in the midst of the speckle noise.\\
Statistical methods have significantly evolved over the last years, and more sophisticated methods based, for instance, upon the receiver operating characteristics (ROC; \citet{2017ApJ...842...14R}) or machine learning \citep{2018A&A...613A..71G} have been proposed.\\  
In our opinion, these techniques can really benefit from the progresses seen in recent years in AO performances and detector technologies, including high performance photon counting detectors, for example the MKID Exoplanet Camera (MEC) recently installed at the SCExAO facility at the Subaru; \citep{2015ESS.....310407M, 2018PASP..130f5001M}. This really makes the difference and turns these methods into a viable, but complementary, strategy to the more commonly used “PSF-subtraction” techniques.\\
In this regard, we note that photon counting sensors may provide uniquely suitable data sets for RQA.
Our analysis demonstrates that the RQA parameters RR, DET, and LAM allow the detection of faint sources in the data. It is worth recalling that the faint sources were injected with a Poisson statistic, thus not all the frames contain photons at the location of the maximum of the PSF of the source itself, as expected in the real case at very high frame rates (i.e. $1$ KHz).\\
In this work we limited ourselves to the analysis of a short sequence of data. The reason for this is primarily due to the computational time required by the RQA method to process longer sequences. However, once more we underline that our purpose here is to explore the possibility of using RQA in astronomical high-contrast imaging. Indeed, the detection of fainter sources would require much longer time series of data to collect enough photons from the faint sources. This is because in the high frame rate regime, the fainter the object the rarer the photons in the data series. \\
However, we note here that achieving a detection contrast of the order of $5 \times 10^{-4}$ by only making use of a sequence of images as short as $2$ s, and without the need of field rotation, is already a good result for at least two reasons. First, achieving such a contrast at visible wavelengths already enables us to investigate several scientific cases. These include, for example, the detection of very-low-mass/sub-stellar companions \citep[e.g.][]{crepp2013,ryu2016,helminiak2016}, the characterization of accreting companions \citep[e.g.][]{close2014,zhou2014,sallum2015}, or the study of the innermost regions of jets from young stellar objects \citep[e.g.][]{antoniucci2016}. Second, the RQA constitutes an independent method, with respect for instance to ADI, SFI, or SFADI, that can be exploited for an independent confirmation of the presence of a specific faint source in the images. Very often, indeed, ADI-like techniques result in a detection map affected by several artifacts. An independent technique based upon a statistical discrimination of the signal of a faint source can be of some help in confirming the detection results. In addition, it is worth noting how the RQA, in contrast to other techniques similar to ADI, does not require field rotation, thus can be applied in those situations where the field rotation itself is not enough to prevent the self-subtraction of the source. \\
Moreover, it is also worth pointing out that, despite the contrast ratios are not extreme compared to those obtained, for example, at infrared wavelengths as we are working in the visible, the RQA, in contrast to other techniques (e.g. SFI), is particularly efficient in the detection of faint sources with angular separations as small as $50$ mas. In addition to this we note that, despite at separations larger than $100$ mas the SFI performs better than RQA in terms of detection contrast, a significant increase of the contrast itself can be obtained by combining the two techniques and by using the RQA as a spatial filter for the residual spatial noise. This is in our opinion one of the most attractive possibilities offered by this technique.\\
Our results show a simultaneous decrease of DET and LAM for faint sources and this reflects an intrinsic change of the underlying dynamics in the phase space due to the presence of a faint additional source.\\ 
Finally, it is interesting to note that the speckle noise is characterized by a larger number of recurrent points in the phase space, if compared to faint sources.\\
One possible explanation is that the arrival of the photons belonging to the faint source in a particular pixel alters the statistics of intensity fluctuations in the pixel itself. This reduces the number of recurrences in the phase space by deviating the trajectories associated with the process represented by the intensity variations of speckles, and diminishing the number of recurrences. Indeed, speckles have a distribution of lifetimes spanning at least two decades, up to $100$ ms \citep{2017JATIS...3b5001S} at visible wavelengths. This imposes a temporal correlation that results in a distinct evolution, in the phase space, of the trajectories associated with the intensity variations of each pixel. The random arrival of the photons of the faint source incident with a Poisson statistic and incident on the same pixels, deviates these trajectories in the phase space reducing the number of recurrences. It is worth noting that the ability of the RQA to disentangle the statistics of the two processes (faint source and speckles) critically depends on the frame rate of the instrument that freezes the evolution of the atmospheric turbulence.  The upcoming LBT instrument SHARK-VIS, by delivering very high frame rate images at least up to $1$ kHz for the brightest sources, will allow us to fully exploit the temporal information contained in the data, and the applied RQA results are an example of this. \\ 

\acknowledgments
The LBT is an international collaboration among institutions in the United States, Italy and Germany. LBT Corporation partners are: Istituto Nazionale di Astrofisica, Italy; The University of Arizona on behalf of the Arizona Board of Regents; LBT Beteiligungsgesellschaft, Germany, representing the Max-Planck Society, The Leibniz Institute for Astrophysics Potsdam, and Heidelberg University; The Ohio State University, and The Research Corporation, on behalf of The University of Notre Dame, University of Minnesota and University of Virginia. SMJ was supported by award FA9550-14-1-0178 from the Air Force Office of Scientific Research. This work was supported by ADONI, the INAF ADaptive Optics National laboratory.

\bibliographystyle{yahapj}

\end{document}